\documentstyle[preprint,tighten,prb,aps,epsf]{revtex}

\def\etal{{\it et al}\ }
\def\Tc{T_{\rm C}}
\def\r{\vec{r}}
\def\uu{^{\mbox{ }}}
\def\s{\sigma}
\def\bs{\vec{\sigma}}
\def\u{\uparrow}
\def\a{\alpha}
\def\l{\lambda}

\def\dd{\downarrow}

\def\en{\epsilon}
\def\las{\langle}
\def\ras{\rangle}
\def\la{\left\las}
\def\lla{\la\la}
\def\llas{\las\las}
\def\ra{\right\ras}
\def\rra{\ra\ra}
\def\rras{\ras\ras}

\def\S{{\bf{S}}}

\def\gmub{g\mu_{\rm B}}

\def\Tr{{\rm Tr}}
\def\I{{\rm I}}
\def\d{{\rm d}}

\begin{document}\draft

\title{Many-body CPA for the Holstein-DE model}
\author{A.C.M.\ Green$^*$}
\address{Department of Mathematics, Imperial College, London SW7 2BZ, UK}
\date{\today}
\maketitle

\begin{abstract}
A many-body coherent potential approximation (CPA) previously developed
for the double exchange (DE) model is extended to include coupling to local
quantum phonons. The Holstein-DE model studied (equal to the Holstein model for
zero Hund coupling) is considered to be a simple model for the colossal
magnetoresistance manganites. We concentrate on effects due to the
quantisation of the phonons, such as the formation of polaron subbands. 
The electronic spectrum and resistivity are
investigated for a range of temperature and electron-phonon coupling strengths.
Good agreement with experiment is found for the Curie temperature and
resistivity with intermediate electron-phonon coupling strength, but phonon
quantisation is found not to have a significant effect in this coupling regime.
\end{abstract}

\pacs{71.10.-w, 71.30.+h, 71.38.+i}

\section{Introduction}\label{sintro}

In this paper we study the Holstein-DE (double exchange) model
\begin{equation}\
H=\sum_{ij\s}t_{ij}\uu c^{\dag}_{i\s}c_{j\s}\uu
-J\sum_i \S_i\uu\cdot\bs_i\uu-h\sum_i L^z_i-g\sum_i n_i\uu\left(b^{\dag}_i
+b_i\uu\right)+\omega\sum_i n^b_i \label{eHDE}
\end{equation}
where $i$ and $j$ are site indices, $c^{\dag}_{i\s}$ and $b^{\dag}_i$
($c_{i\s}\uu$ and $b_i\uu$) create (annihilate) an electron of
spin $\s$ and a phonon respectively, $\S_i$ is a local spin,
$\bs_i=(1/2)\sum_{\s\s'}c^{\dag}_{i\s}
\vec{\s}_{\s\s'}\uu c_{i\s'}\uu$ is
the electron spin operator ($\vec{\s}_{\s\s'}$ being the Pauli matrix),
$L^z_i=S^z_i+\s^z_i$ is the $z$-component of angular momentum,
$n_i\uu=\sum_{\s}c^{\dag}_{i\s}c_{i\s}\uu$ and $n^b_i=b^{\dag}_i b_i\uu$.
The parameter $t_{ij}$ is the hopping integral, $J>0$ is the
Hund coupling, $h$ is the Zeeman energy, $g$ is the electron-phonon
coupling strength and $\omega$ is the Einstein phonon energy. $H$ is a
model for the colossal magnetoresistance (CMR) manganite compounds with
the double degeneracy of the conduction band neglected and a simplified
form assumed for the electron-phonon coupling and phonon
dispersion. The electron-phonon
coupling in Eq.\ (\ref{eHDE}) is of the breathing-mode form, i.e.\
$-g'\sum_i n_i x_i$ in the classical limit (where $x_i$ is the phonon
displacement), but we regard it as an effective Jahn-Teller coupling.

Hamiltonian (\ref{eHDE})
was first studied by R\"oder \etal \cite{rRoder},
who treated the Hund coupling using a mean-field approximation and the
electron-phonon coupling using a variational Lang-Firsov approximation.
The same authors later used a similar method to study a more realistic
model for CMR systems \cite{rRoder2}. In this paper we treat both the Hund 
and the electron-phonon coupling using an
extension of a many-body coherent potential approximation (CPA) previously 
derived for the DE model \cite{rUs1,rUs2}. The CPA treats the Hund
coupling better than mean-field theory and has the advantage over
Lang-Firsov variational methods that the whole of the electronic
spectrum can be studied, not just the coherent polaron band near the Fermi 
energy. In the limit of classical spins and phonons Millis \etal used 
dynamical mean-field theory (DMFT) to study another more realistic model
for CMR materials \cite{rMillisII,rMillis}. Here however we concentrate on 
the effects of quantisation of the phonons.

Our approach has many similarities with studies of the Holstein
model using DMFT, which have been carried
out for the classical phonon \cite{rMillisI} and empty-band \cite{rCiuchi} 
limits in which the model is a one-electron problem. Indeed the standard
dynamical CPA is equivalent to DMFT for one-electron problems such as the
binary alloy \cite{rGeorges}, the DE and Holstein models
in the empty-band limit \cite{rCiuchi,rKubo,rSumi}, and the DE model with
classical local spins \cite{rUs2}. DMFT should be regarded as the correct
extension of the CPA to many-body problems \cite{rJanis}.
For the current many-body problem we regard our
CPA as an approximate solution of DMFT, or as an extrapolation from the
one-electron case. The CPA has the advantage of
relative analytic simplicity, but does not treat the many-body dynamics as
well as DMFT, retaining too much one-electron character.

The many-body CPA derived for the finite $S$ DE model in Ref.\
\onlinecite{rUs1} and Ref.\ \onlinecite{rUs2}
was based on Hubbard's scattering correction approximation for 
the Hubbard model \cite{rHubbardIII}. Hubbard's approximation was derived
by decoupling Green function equations of motion (EOM) according to an alloy
analogy in which electrons of one spin are frozen while the propagation of 
those of the opposite spin is considered (within the CPA). Although more
modern formulations of the CPA exist Hubbard's EOM approach was found to be
particularly suitable for extension to the DE model, where the possibility of
electrons exchanging (spin) angular momentum with local spins complicates the
problem. The resulting many-body CPA was exact in the atomic limit
$t_{ij}\rightarrow 0$ and recovered the one-electron
CPA/DMFT in the empty-band and classical spin ($S\rightarrow\infty$) limits. 
 
In Sec.\ \ref{sAL} we solve the atomic limit of Hamiltonian (\ref{eHDE}),
and in 
Sec.\ \ref{sGF} we extend our many-body CPA to the Holstein-DE model. The
properties of the CPA solution are discussed in Sec.\ \ref{sResults}, and the
special case ($J=h=0$) of the Holstein model is considered in
Sec.\  \ref{sHolstein}. We give a summary in Sec.\  \ref{sSummary}.

\section{The atomic limit}\label{sAL}

In the atomic limit $t_{ij}\rightarrow 0$ $H$ is exactly solvable using
the canonical transformation $H\mapsto
\tilde{H}=e^{\nu^{\dag}}H e^{\nu}$ where $\nu=\frac{g}{\omega}
n(b^{\dag}-b)$ (Ref.\ \onlinecite{rLangFirsov}) (we drop site indices).
In the presence of electron-phonon coupling the
phonon potential is of the displaced harmonic oscillator form, and the
effect of the canonical transformation is to shift the operators
to take account of this: $b\mapsto b+\frac{g}{\omega}n$ and
$c_{\s}\uu\mapsto Xc_{\s}\uu$ where $X=\exp(g/\omega(b^{\dag}-b))$.
This transformation decouples the Hamiltonian $\tilde{H}=H_b+H_f$
into a bosonic component $H_b=\omega n^b$ and a fermionic component
$H_f=-J\S\cdot\bs-hL^z-g^2/\omega\,n^2$
where $g^2/\omega=:\lambda\omega$ is the binding energy of a polaron.

The one-electron Green function $G_{\s}(t)=-i\theta(t)
\las \{c_{\s}(t),c_{\s}^{\dag}\}\ras$ can be separated into fermionic and
bosonic traces using the invariance of the trace under cyclic
permutations and $e^{\nu^{\dag}}e^{\nu}=1$,
\begin{eqnarray}\nonumber
G_{\s}(t)=-i\theta(t)\left[
\frac{\Tr_f\{e^{-\beta H_f}e^{iH_f t}c_{\s}e^{-iH_f t}c^{\dag}_{\s}\}}
{\Tr_f\{e^{-\beta H_f}\}}
\frac{\Tr_b\{e^{-\beta H_b}e^{iH_b t}X e^{-iH_b t}X^{\dag}\}}
{\Tr_b\{e^{-\beta H_b}\}}\right.\\\nopagebreak\left.+
\frac{\Tr_f\{e^{-\beta H_f}c^{\dag}_{\s}e^{iH_f t}c_{\s}e^{-iH_f t}\}}
{\Tr_f\{e^{-\beta H_f}\}}
\frac{\Tr_b\{e^{-\beta H_b}X^{\dag}e^{iH_b t}X e^{-iH_b t}\}}
{\Tr_b\{e^{-\beta H_b}\}}\right].
\end{eqnarray}
We evaluate the bosonic traces directly and the fermionic traces using the
equation of motion (EOM) method, and in the energy representation
$G_{\s}(\en)=\int_{-\infty}^{\infty}\d t\,e^{i\en t}G_{\s}(t)$ obtain
\begin{eqnarray}\nonumber
G_{\s}(\en)=\sum_{r=-\infty}^{\infty}\frac{\I_r
\left(2\l\sqrt{b(\omega)(b(\omega)+1)}\,\right)}
{(2S+1)\exp(\l(2b(\omega)+1))}\sum_{\alpha=\pm}\left[
\frac{e^{r\beta\omega/2}\,W^{0+}_{\alpha\s}
+e^{-r\beta\omega/2}\,W^{0-}_{\alpha\s}}{
\en-\alpha JS/2+h\s/2+\lambda\omega(1+2\delta_{\alpha+})+\omega r}\right.\\
\left.+\frac{e^{r\beta\omega/2}\,W^{1+}_{\alpha\s}
+e^{-r\beta\omega/2}\,W^{1-}_{\alpha\s}}
{\en+\alpha J/2(S+1)+h\s/2+\lambda\omega(1+2\delta_{\alpha+}) +\omega r}
\right].\label{eGAL0}
\end{eqnarray}
Here the weight factors
\begin{mathletters}
\begin{eqnarray}
W^{0\gamma}_{\alpha\s}= (S+1)\la n^{\alpha}_{-\s}n^{\gamma}_{\s}\ra
-\alpha\s \la S^z n^{\alpha}_{-\s}n^{\gamma}_{\s}\ra
 +\left(1-\delta_{\alpha\gamma}\right)\la S^{-\s}\s^{+\s}\ra\\
W^{1\gamma}_{\alpha\s}= S\la n^{\alpha}_{-\s}n^{\gamma}_{\s}\ra+
\alpha\s\la S^z n^{\alpha}_{-\s}n^{\gamma}_{\s}\ra -
\left(1-\delta_{\alpha\gamma}\right)\la S^{-\s}\s^{+\s}\ra,
\end{eqnarray}
\end{mathletters}
$\alpha$, $\gamma=\pm$, $\delta_{\alpha+}=1$ for $\alpha=+$ and $0$ for
$\alpha=-$,
$\I_r$ is the modified Bessel function, 
$b(\omega)=1/(\exp(\beta\omega)-1)$ is the Bose function, and we define
$n^+_{\s}=n_{\s}$, $n^-_{\s}=1-n_{\s}$, and $S^{-\s}\s^{+\s}=S^-\s^+$ for
$\s=\u$ and $S^+\s^-$ for $\s=\downarrow$.

\begin{figure}[h]
\begin{center}
\leavevmode
\hbox{%
\epsfxsize=0.6\textwidth
\epsffile{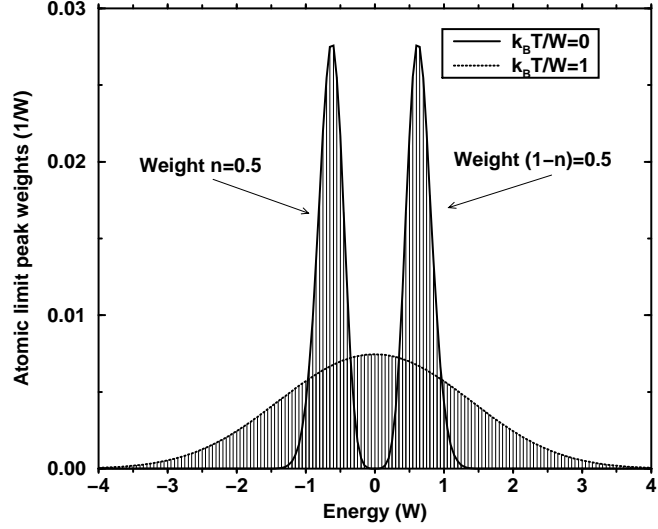}
}
\end{center}
\caption{Peak weights of the atomic limit spectrum at low and high temperature.
Plot is for the (forced) paramagnetic state with $S=J=\infty$, $h=0$, $n=0.5$,
$\omega/W=0.05$ and $g/W=0.18$, $W$ being an energy parameter.
\label{fALGF}}
\end{figure}

In the ($J=h=0$) case of the Holstein model Eq.\ (\ref{eGAL0}) reduces to the
formula
\begin{eqnarray}\nonumber
G_{\s}(\en)=\sum_{r=-\infty}^{\infty}\frac{\I_r
\left(2\l\sqrt{b(\omega)(b(\omega)+1)}\,\right)}
{\exp(\l(2b(\omega)+1))}\left[
\frac{e^{r\beta\omega/2}\las n^-_{-\s}n_{\s}\ras+
e^{-r\beta\omega/2}\las n^-_{-\s}n^-_{\s}\ras}
{\en+\lambda\omega+\omega r}\right.\\ \left.
+\frac{e^{r\beta\omega/2}\las n_{-\s}n_{\s}\ras+
e^{-r\beta\omega/2}\las n_{-\s}n^-_{\s}\ras}
{\en+3\lambda\omega+\omega r}\right],\label{eHolsteinAL}
\end{eqnarray}
but we are mostly interested in the strong Hund-coupling limit,
so we shift the energy to have the zero near
the Fermi level, $\en\mapsto\en-JS/2$, and let
$J\rightarrow\infty$. In this limit we find 
\begin{equation}\label{eALG}
G_{\s}(\en)=\sum_{r=-\infty}^{\infty}\frac{\I_r
\left(2\l\sqrt{b(\omega)(b(\omega)+1)}\,\right)}
{(2S+1)\exp(\l(2b(\omega)+1))}\frac{
e^{r\beta\omega/2}\, W^+_{\s}
+e^{-r\beta\omega/2}\,W^-_{\s}}{\en+h\s/2+\omega r+\lambda\omega}
\end{equation}
where the weight factors
\begin{mathletters}\begin{eqnarray}
 W^+_{\s}=& \la(S+1+\s S^z)n_{\s}+S^{-\s}\s^{+\s}\ra&=
(2S+1)\left(\frac{n}{2}+\s\la\s^z\ra\right)\\
 W^-_{\s}=&\la(1-n)(S+1+\s S^z)\ra&=
 (S+1)(1-n)+\s\la S^z\ra-2S\s\la\s^z\ra.
\end{eqnarray}\end{mathletters}
The paramagnetic state spectrum of Eq.\ (\ref{eALG})
is plotted in Fig.\ \ref{fALGF}
for the classical spin limit $S\rightarrow\infty$ at quarter-filling
$n=0.5$. The spectrum consists of delta-function peaks separated in energy by
$\omega$, and for clarity we include the peaks' envelope curve
in Fig.\  \ref{fALGF}. Note that the symmetry of the spectrum about zero
energy is due to choice of filling $n=0.5$; in general the lower and upper
(zero temperature) `bands' have weights $n$ and $1-n$ respectively.
By counting weights it may be seen that for any $n$ the zero temperature
chemical potential $\mu(T=0)=0$ lies in the zero energy peak in the middle of
the pseudogap.

\section{The CPA Green function}\label{sGF}

We now derive a many-body CPA for the one-electron Green function
$G^{ij}_{\s}(\en)$ of the full Hamiltonian (\ref{eHDE}). As discussed in
the introduction we proceed by decoupling equations of motion (EOM),
adapting decoupling approximations previously used for the DE model
\cite{rUs2}. Recall that with the fermionic definition of Green
functions, $\llas A\,;C\rras_{\en}=-i\int_0^{\infty}\d t
\,\exp(i\en t)\las\{A(t),C\}\ras$, the EOM is
\begin{equation}
\en\,\lla A\,;C\rra_{\en}=\la\{A,C\}\ra+\lla[A,H]\,;C\rra_{\en}.
\end{equation}
As in Ref.\ \onlinecite{rUs2}
and in the original version of this method due to Hubbard
\cite{rHubbardIII} we split the Green function into a low-energy component
$G^{ij-}_{\s}(\en)=\llas n^-_i c_{i\s}\uu;c^{\dag}_{j\s}\rras_{\en}$ and a
high-energy component 
$G^{ij+}_{\s}(\en)=\llas n^+_i c_{i\s}\uu;c^{\dag}_{j\s}\rras_{\en}$, i.e.\
$G^{ij}_{\s}(\en)=\sum_{\alpha=\pm}G^{ij\alpha}_{\s}(\en)$. To close the
system of EOM we in fact need to introduce the Green functions
\begin{mathletters}
\begin{eqnarray}\label{edefS}
S^{ij\a}_{\s}(\r,\en)&=&
\lla \Gamma_i(\r) n^{\a}_i c_{i\s}\uu;
c^{\dag}_{j\s}\rra_{\en}\\
T^{ij\a}_{\s}(\r,\en)&=&
\lla \Gamma_i(\r)
n^{\a}_i S^{-\s}_i c_{i-\s}\uu;c^{\dag}_{j\s}\rra_{\en}
\end{eqnarray}
\end{mathletters}
where $\r=(\rho,\phi,\theta)$ is a parameter and the operator 
$\Gamma_i(\r)=\exp(\rho S^z_i)\exp(\phi b^{\dag}_i)\exp(\theta b_i)$.
Note that $\Gamma$ is a generating function for the operators
$S^z$, $b^{\dag}$ and $b$, so that $\partial^n/\partial\phi^n\Gamma_i(\r)
=(b^{\dag}_i)^n\Gamma_i(\r)$ for instance.
This is convenient in allowing us to close
the system of EOM with a minimal number of equations.

When writing the EOM we use the convenient commutation identities
$[e^{\phi b^{\dag}},b]=-\phi e^{\phi b^{\dag}}$ and
$[e^{\theta b},b^{\dag}]=\theta e^{\theta b}$, and the Feynman
operator disentanglement relation
$e^{A+B}=e^A e^B e^{-1/2[A,B]}$, which holds if
$[[A,B],A]=[[A,B],B]=0$. We also work for $\s=\u$; the $\s=\dd$ equations
can be obtained using the symmetry of $H$.
We introduce the operator
\begin{equation}
K^{\alpha}(\theta,\phi)=\omega\left(\phi\frac{\partial}{\partial\phi}
-\theta\frac{\partial}{\partial\theta}\right)
+g\left(\frac{\partial}{\partial\phi}+\frac{\partial}{\partial\theta}
+\theta+\delta_{\alpha+}(\theta-\phi)\right),
\end{equation}
and obtain the (exact) EOM
\begin{eqnarray}\nonumber
\left[\en+\frac{h}{2}+\frac{J}{2}\frac{\partial}{\partial\rho}+
K^{\alpha}(\theta,\phi)\right]S^{ij\a}_{\u}(\r,\en)+
\frac{J}{2}e^{\rho\delta_{\alpha+}}T^{ij\a}_{\u}(\r,\en)=\delta_{ij}
\la\Gamma(\r)n^{\alpha}_{\dd}\ra\hskip 1in
\\ +\lla\Gamma_i(\r)\left[n^{\alpha}_{i\dd},H_0\right]c_{i\u};c^{\dag}_{j\u}
\rra_{\en}
+\sum_k t_{ik}\lla\Gamma_i(\r)n^{\alpha}_{i\dd}c_{k\u};c^{\dag}_{j\u}
\rra_{\en}\label{eEOMS1}
\end{eqnarray}
for $S^{ij\a}_{\u}(\r,\en)$ and
\begin{eqnarray}\nonumber
\left[\en+\frac{h}{2}-\frac{J}{2}
\left(\delta_{\alpha-}+\frac{\partial}{\partial\rho}\right)+
K^{\alpha}(\theta,\phi)\right]T^{ij\a}_{\u}(\r,\en) \hskip 2.5in\\ \nonumber+
\frac{J}{2}e^{-\rho\delta_{\alpha+}}
\left[S(S+1)+\alpha\frac{\partial}{\partial\rho}-
\frac{\partial^2}{\partial\rho^2}\right]S^{ij\a}_{\u}(\r,\en)=
-\alpha\delta_{ij}\la\Gamma(\r)S^-\s^+\ra\\
+
\lla\Gamma_i(\r)\left[n^{\alpha}_{i\u},H_0\right]S^-_i c_{i\dd};
c^{\dag}_{j\u}
\rra_{\en}
+\sum_k t_{ik}\lla\Gamma_i(\r)n^{\alpha}_{i\u}S^-_i c_{k\dd};c^{\dag}_{j\u}
\rra_{\en}\label{eEOMT1}
\end{eqnarray}
for $T^{ij\a}_{\u}(\r,\en)$. Here $H_0$ is the electron hopping term of
the Hamiltonian and we have dropped site indices in the expectations,
assuming a homogeneous state. Equations (\ref{eEOMS1}) and (\ref{eEOMT1})
should be compared with their analogues for the DE model case: equations
(4) and (5) in Ref.\ \onlinecite{rUs2}.

As usual we now neglect the penultimate Green functions in equations
(\ref{eEOMS1}) and (\ref{eEOMT1}) (the ones containing
$[n^{\alpha}_{i\s},H_0]$). This corresponds to making the alloy analogy.
We treat the final Green functions in these equations using a CPA and
treat all other terms exactly. In fact we use the approximations
\begin{mathletters}
\begin{eqnarray}
\sum_k t_{ik} \lla\Gamma_i(\r) n^{\alpha}_{i\dd}
c_{k\u}\uu\,;c^{\dag}_{j\u}\rra_{\en}\approx
\la\Gamma(\r) n^{\alpha}_{\dd}\ra\left(
\sum_k t_{ik}G^{kj}_{\u}(\en)-J_{\u}(\en)G^{ij}_{\u}(\en)\right)
+J_{\u}(\en)S^{ij\a}_{\u}(\r,\en) \label{eSC1}\\ \nonumber
\sum_k t_{ik}\lla \Gamma_i(\r) n^{\alpha}_{i\u} S^-_i
c_{k\dd}\uu\,;c^{\dag}_{j\u}\rra_{\en}\approx
-\alpha\la\Gamma(\r) S^-\s^+\ra\left(
\sum_k t_{ik}G^{kj}_{\u}(\en)-J_{\u}(\en)G^{ij}_{\u}(\en)\right)\hskip 0.45in\\
+J_{\dd}(\en+h)T^{ij\a}_{\u}(\r,\en)\label{eSC2}
\end{eqnarray}
\end{mathletters}
where $J_{\s}(\en)=\en-\Sigma_{\s}(\en)-G_{\s}(\en)^{-1}$,
$\Sigma_{\s}(\en)$ and $G_{\s}(\en)$ being the self-energy and local Green
function respectively. The function $E_{\s}(\en)=\en-J_{\s}(\en)$
is related to the Weiss function of
DMFT, $E_{\s}(i\omega_n)\equiv{\cal G}^{-1}_{\s}(i\omega_n)$, $i\omega_n$ being
a fermionic Matsubara frequency.
Equations (\ref{eSC1}) and (\ref{eSC2}) are
generalisations of Hubbard's scattering correction approximation
\cite{rHubbardIII}, and lead to the CPA equations in the case of the DE
model. We make these particular approximations since Eq.\ (\ref{eSC1}) and
Eq. (\ref{eSC2}) are of the usual CPA form, but do not give a formal
justification. 

We define $E^h_{\s}(\en)=E_{\s}(\en+h\delta_{\s\dd})+\s h/2$ and
$\lambda^{ij}_{\s}(\en)=\delta_{ij}+\sum_k t_{ik}G^{kj}_{\s}(\en)
-J_{\s}(\en)G^{ij}_{\s}(\en)$, and from equations (\ref{eEOMS1}),
(\ref{eEOMT1}), (\ref{eSC1}) and (\ref{eSC2}) obtain
\begin{eqnarray}
\nonumber
\left[\begin{array}{cc}
E^h_{\u}(\en)+\frac{J}{2}\frac{\partial}{\partial\rho}+
K^{\alpha}(\theta,\phi) &
\frac{J}{2}e^{\rho\delta_{\alpha+}}\\
\frac{J}{2}e^{-\rho\delta_{\alpha+}}
\left(S(S+1)+\alpha\frac{\partial}{\partial\rho}-
\frac{\partial^2}{\partial\rho^2}\right) &
E^h_{\dd}(\en)-\frac{J}{2}
\left(\delta_{\alpha-}+\frac{\partial}{\partial\rho}\right)+
K^{\alpha}(\theta,\phi)
\end{array}\right]
\left(\begin{array}{c}
S^{ij\a}_{\u}(\r,\en)\\
T^{ij\a}_{\u}(\r,\en)
\end{array}\right)\\
\approx \lambda^{ij}_{\u}(\en)
\left(\begin{array}{c}
\la\Gamma(\r)n^{\alpha}_{\dd}\ra\\
-\alpha\la\Gamma(\r)S^-\s^+\ra
\end{array}\right).\label{eCPA}
\end{eqnarray}
We make no further approximations. We use the top row of Eq. (\ref{eCPA}) to
eliminate $T^{ij\a}_{\u}(\r,\en)$, obtaining a second-order linear
(parabolic) PDE for $S^{ij\a}_{\u}(\r,\en)$. We take $i=j$ and use
$\lambda^{ii}_{\s}(\en)=1$ (Ref.\ \onlinecite{rUs2}).
In the strong Hund-coupling
limit $J\rightarrow\infty$, which we take with the energy origin shifted
as $\en\mapsto\en-JS/2$, this second-order PDE
simplifies to the first-order linear PDE
\begin{eqnarray}\nonumber
\left[\frac{(1+S)E^h_{\u}(\en)+SE^h_{\dd}(\en)}{2S+1}
+\frac{E^h_{\u}(\en)-E^h_{\dd}(\en)}{2S+1}\frac{\partial}{\partial\rho}
+K(\theta,\phi)\right]S_{\u}(\r,\en)=\hskip 1in\\
\la\Gamma(\r)\frac{(S+1+S^z)n^-_{\dd}+S^-\s^+}{2S+1}\ra.\label{ePDE1}
\end{eqnarray}
Note that in this limit we may assume $\alpha=-$. We change $(\theta,\phi)$
variables to $\Phi=(\theta-g/\omega)(\phi+g/\omega)$ and
$\Theta=1/\omega\ln(\phi+g/\omega)$, in terms of which
\begin{equation}
K(\Theta,\Phi)=g\left(\frac{g}{\omega}+\Phi e^{-\omega\Theta}\right)+
\frac{\partial}{\partial\Theta}.
\end{equation}
In the new $(\rho, \Theta, \Phi)$ system of variables Eq.\ (\ref{ePDE1})
contains
derivatives with respect to $\rho$ and $\Theta$ only, facilitating its
solution. We find (see appendix) for $\rho=\theta=\phi=0$
\begin{eqnarray}\nonumber
G_{\u}(\en)=\sum_{m,n=0}^{\infty}\frac{(-1)^n \l^{(m+n)/2}}
{m!\,n!\,e^{\l}} \hskip 4in\\ \hskip -0.25in\times
\la
\frac{e^{-g/\omega b^{\dag}}(b^{\dag})^m(b-g/\omega)^n e^{g/\omega b}
[(S+1+S^z)n^-_{\dd}+S^-\s^+]}
{(S+1+S^z)E^h_{\u}(\en)+(S-S^z)E^h_{\dd}(\en)+(2S+1)(\l\omega+(m-n)\omega)}
\ra \label{eG0}
\end{eqnarray}
where $\la\ra$ denotes quantum and statistical averaging and
$S^z$ in the denominator acts on the left.

In principle the average in Eq.\ (\ref{eG0})
should be determined self-consistently,
but this is difficult to carry out. In previous many-body CPAs, e.g.\ Hubbard's
for the Hubbard model and ours for the DE model, it was found that
the hopping does not affect the total weight in a band near a given atomic
limit peak, at least when the bands are separated so that the
weight associated with a given atomic limit peak is a meaningful quantity.
For simplicity we therefore assume that all averages take their atomic limit
values. Note that owing to the degeneracy of the atomic limit states this
says nothing about the spin polarisation. This assumption means that we cannot 
take account of the effects of electron hopping on the phonon distribution.

The right-hand side of Eq.\ (\ref{eG0})
then depends on the half-bandwidth $W$
only through $E^h_{\s}(\en)$. We change summation variables $(m,n)$ to
$r=m-n$ and $s=m+n$ and use local spin projection operators $P(S^z=m^z)$ to
pull the denominator of Eq.\ (\ref{eG0}) out of the average. Since
$\lim_{W\rightarrow 0}[(S+1+m^z)E^h_{\u}(\en)+(S-m^z)E^h_{\dd}(\en)]=
(2S+1)(\en+h/2)$ we can match our averages (summed over $s$) with atomic limit 
peak weights to see that
\begin{eqnarray}\nonumber
G_{\u}(\en)=\sum_{r=-\infty}^{\infty}
\frac{\I_r\left(2\l\sqrt{b(\omega)(b(\omega)+1)}\,\right)}
{\exp(\l(2b(\omega)+1))} \hskip 3.5in\\
\times\la\frac{e^{-r\beta\omega/2}(1-n)(S+1+S^z)+ e^{r\beta\omega/2}
[(S+1+S^z)n_{\u}+S^-\s^+]}{
(S+1+S^z)E^h_{\u}(\en)+(S-S^z)E^h_{\dd}(\en)+(2S+1)(\l\omega+r\omega)}\ra.
\label{eG1}
\end{eqnarray}
This should be compared with the atomic limit expression Eq.\ (\ref{eALG}),
to which Eq.\ (\ref{eG1}) reduces as $W\rightarrow 0$.

Now in the case of the
empty-band limit of the Holstein model \cite{rSumi,rCiuchi} one is used to
obtaining a CPA/DMFT expression for $G_{\s}$ in the form of a continued
fraction. In Eq.\ (\ref{eG1}) we have a simpler expression involving a
sum over the atomic limit peaks, despite the more complex nature of the 
problem which we are considering (the many-electron case with both Holstein and
DE interactions). One might suspect that our CPA is cruder
than the one-electron CPA. In fact our expression for $G_{\u}$ for the
Holstein model (given in Sec.\  \ref{sHolstein}) in the limit
$n\rightarrow 0$ is equivalent to making the approximation
\begin{equation}\label{ebadapprox}
E^h_{\s}(\en+r\omega)\approx E^h_{\s}(\en)+r\omega
\end{equation}
in the one-electron CPA expression. We will mainly consider the case of
an elliptic bare density of states (DOS), 
$D(\en)=2/(\pi W^2)\sqrt{W^2-\en^2}$, for which it be shown that
$E(\en)=\en-W^2 G(\en)/4$. For the elliptic DOS approximation
Eq.\ (\ref{ebadapprox}) is thus equivalent to neglecting
energy shifts in the Green function on the right-hand side of the CPA
equation. Since we do not recover the one-electron CPA/DMFT as $n\rightarrow
0$, unlike in the case of the bare DE model \cite{rUs2},
our CPA for the Holstein-DE model is not as good as our CPA for the DE model. 
We choose however to accept the increased crudeness of the
approximation in return for the greatly increased simplicity; a CPA which
correctly reduced to the one-electron CPA as $n\rightarrow 0$ would
probably be analytically intractable in the many-body case.

\subsection{Calculation of Curie temperature}

Using mean-field arguments Millis \etal \cite{rMillisa} claimed that the bare
DE model predicts a Curie temperature $\Tc$ at least an order of magnitude too
large. However, subsequent more reliable treatments of the DE model taking into
account quantum fluctuations
\cite{rFurukawaTc,rUs2} showed that the DE model's $\Tc$ is in fact in
reasonable agreement with experiment. Now as discussed by R\"oder \etal
\cite{rRoder} phonon coupling suppresses $\Tc$. We therefore calculate $\Tc$
to see if a phonon coupling strength $g$ exists that gives a much
larger resistivity (than the $g=0$ case) without making $\Tc$ unphysically
small.

For simplicity we work in the classical limit $S=\infty$, in which
$E^h_{\s}(\en)=E_{\s}(\en)$ since $h\sim 1/S$, and specialise to
the case of an elliptic bare DOS, where as mentioned above
$E_{\s}(\en)=\en-W^2 G_{\s}(\en)/4$ is just a function of $G_{\s}$.
We set $h=0$ in Eq.\ (\ref{eG1}) and expand
$G_{\s}$ about the paramagnetic state to 
first order (in $\delta\las\s^z\ras$ or $\delta\las S^z\ras$), obtaining
\begin{equation}
\delta G_{\u}(\en)=\left(\sum_r w_r(\en)
e^{-r\beta\omega/2}\right)\delta\la S^z\ra
+4\left(\sum_r w_r(\en)\sinh(r\beta\omega/2)\right)\delta\la\s^z\ra\label{edG}
\end{equation}
where
\begin{equation}
w_r(\en)= \frac{\I_r\left(2\lambda\sqrt{b(\omega)(1+b(\omega))}\,\right)}
{2\exp(\lambda(1+2b(\omega)))}
\frac{(E(\en)+\omega r)^{-1}}{1+(W^2/12)G'(\en)/E'(\en)},
\end{equation}
$A'(\en)=\d A(\en)/\d\en$ and we drop spin indices on paramagnetic state
quantities. From the spectral theorem
$\la\s^z\ra={\cal I}[G_{\u}-G_{\dd}]/2$ where
\begin{equation}
{\cal I}[A]=\int_{-\infty}^{\infty}\frac{\d\en}{\pi}f(\en-\mu)
{\rm Im}\left[A(\en-i0)\right],
\end{equation}
$f(\en-\mu)$ being the Fermi function. Applying $\cal I$ to Eq.\ (\ref{edG})
and
rearranging leads to
\begin{equation}
\delta\la\s^z\ra=\frac{\sum_r {\cal I}[w_r]\exp(-r\beta\omega/2)}
{1-4\sum_r {\cal I}[w_r]\sinh(r\beta\omega/2)}\delta\la S^z\ra.\label{eds}
\end{equation}

In Ref.\ \onlinecite{rUs2} we showed (in the DE model case) that the CPA for
electronic Green functions does not give a good estimate of local spin 
expectations. Fortunately for $S=\infty$ we can use DMFT to obtain an
expression for $\delta\las S^z\ras$ in terms of $\delta G_{\u}$ which is exact
in the infinite-dimensional limit. After integrating out the bosonic degrees
of freedom the DMFT effective action can be written in the Matsubara
representation as
\begin{eqnarray}\nonumber
\tilde{S}=-\sum_n
\left(\begin{array}{cc} c^{\dag}_{n\u} & c^{\dag}_{n\dd}\end{array}\right)
\left(\begin{array}{cc} E_{\u}(i\omega_n)+(J/2) S^z & (J/2) S^- \\
(J/2) S^+ & E_{\dd}(i\omega_n)-(J/2) S^z \end{array}\right)
\left(\begin{array}{c} c_{n\u} \\ c_{n\dd} \end{array}\right)-\beta h S^z\\
+\int_0^{\beta}\d\tau\int_0^{\beta}\d\tau'\,n_i(\tau){\tilde U}(\tau-\tau')
n_i(\tau')
\label{etS}
\end{eqnarray}
where we work for $S=\infty$, $h$ and $J$ finite, and the $n$
subscripts refer to fermionic Matsubara frequencies.
The last term in Eq.\ (\ref{etS})
is an attractive Hubbard-like term, with the
Fourier transform of the interaction given by
${\tilde U}(i\omega_n)=-(1/2)g^2/(\omega^2+\omega_n^2)$.
It is retarded in imaginary time and
originates from the phonon coupling \cite{rGeorges}. We expand $\tilde{S}$
about the $h=0$ paramagnetic state with action $\tilde{S}_0$
and partition function $Z_0$,
$\tilde{S}=\tilde{S}_0+\delta{\tilde{S}}+\cdots$ where
\begin{equation}
\delta\tilde{S}=-\beta hS^z-\sum_n\left(\delta E_{\u}(i\omega_n)c^{\dag}_{n\u}
c_{n\u}\uu+\delta E_{\dd}(i\omega_n) c^{\dag}_{n\dd}c_{n\dd}\uu\right),
\end{equation}
and in terms of $\delta\tilde{S}$ we have
\begin{equation}
\delta\la S^z\ra=-\frac{1}{Z_0}\int\d^2 S\left(\prod_{n\s}\d c^{\dag}_{n\s}
\d c_{n\s}\right)S^z e^{-\tilde{S}_0}\delta\tilde{S}\label{edSz}
\end{equation}
where 
$\int\d^2 S$ is the integral over the surface of the unit sphere.

Now $\delta E_{\s}(i\omega_n)=-W^2\delta G_{\s}(i\omega_n)/4$ for an
elliptic DOS, and we use the relation $\beta^{-1}\sum_n g(i\omega_n)=
{\cal I}[g]$, which holds for functions $g$ analytic off the real axis, to
write Eq.\ (\ref{edSz}) as
\begin{equation}
\delta\la S^z\ra=
\beta\left\{\frac{h}{3}-\frac{W^2}{2}{\cal I}
\left[\left(\frac{\partial S(\r,\en)}{\partial\rho}\right)_{\r=0}
\delta G_{\u}(\en)\right]
\right\}.\label{edSz2}
\end{equation}
Note that the effects of phonon coupling enter only implicitly via
Green functions. This is expected as the electron-phonon coupling is
spin-symmetric. Setting $h=0$ in Eq.\ (\ref{edSz2}) and using (\ref{edG}),
(\ref{eds})
and $\partial S(\r,\en)/\partial\rho\vert_{\r=0}=G(\en)/3$
we obtain the Curie temperature equation
\begin{equation}
k_{\rm B}\Tc=-\frac{W^2}{6}\left[\frac{\sum_r{\cal I}[Gw_r]e^{-r\beta\omega/2}
+4\sum_{rs}{\cal I}[Gw_r]{\cal I}[w_s]\sinh(\beta\omega(r-s)/2)}
{1-4\sum_r{\cal I}[w_r]\sinh(r\beta\omega/2)}\right]\label{Tceqn}
\end{equation}
upon dividing by $\delta\las S^z\ras$.
We discuss the value of $\Tc$ in the next section.

\section{Results}\label{sResults}

We now discuss numerical results obtained using our CPA, for
simplicity using the elliptic bare DOS and working at
$J=S=\infty$ and $n=0.5$.
In the spin-saturated state the minority-spin weight at low
energy is of order $1/S$, so the classical limit $S\rightarrow\infty$ is
convenient as we do not need band-shifts, which are difficult to obtain within
the CPA, for consistency of the saturated ferromagnetic state. Quarter-filling
$n=0.5$ is used because owing to the symmetry of the spectrum about zero energy
the chemical potential $\mu(n=0.5)=0$ for all $T$. For a
homogeneous state the doping has a qualitative effect only as
$n\rightarrow 0$ or 1, and the results for a more physical value
$n\approx 0.7$ are similar in form to those at $n=0.5$. Note however that the
quantitative predictions of the model are very sensitive to the model
parameters, especially the electron-phonon coupling strength but also the
doping, so if the physical doping value is used the model parameters must be
adjusted to retain quantitative agreement with experiment. We take
$\omega/k_{\rm B}=0.05W/k_{\rm B}\sim 600$K for $W\sim 1$eV. Zhao \etal report
that $\omega/k_{\rm B}\sim 100$K for La$_{1-x}$Ca$_x$MnO$_3$
(Ref.\ \onlinecite{rZhao}) so this may be a bit large.

\begin{figure}[h]
\begin{center}
\leavevmode
\hbox{%
\epsfxsize=0.6\textwidth
\epsffile{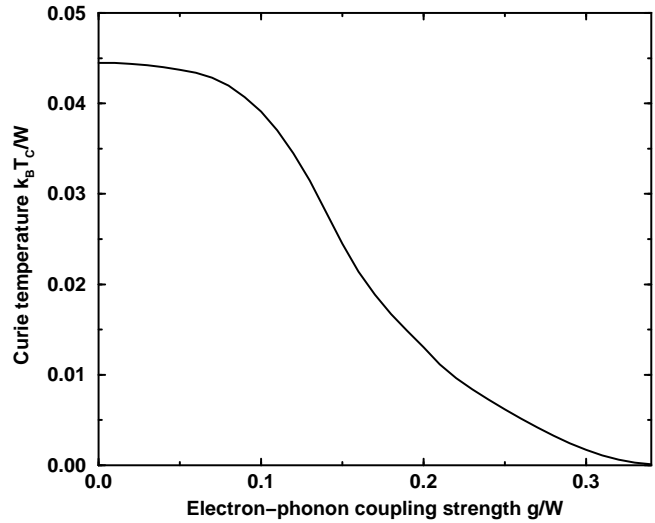}
}
\end{center}
\caption{Suppression of the Curie temperature $\Tc$ with increasing
electron-phonon coupling $g/W$. Plot is for $S=J=\infty$, $h=0$, $n=0.5$ and
$\omega/W=0.05$.\label{fTc}}
\end{figure}

The Curie temperature $\Tc$ obtained from equations
(\ref{Tceqn}) and (\ref{eG1}) is
plotted against electron-phonon coupling strength $g$ in Fig.\  \ref{fTc}.
It will
be found later that $g\approx 0.16W$ gives reasonable values for the
resistivity. For values of $g$ in this range $\Tc$ is only suppressed by about 
a factor $\sim 2$, so for $W\sim$1eV is still compatible with experiment. 

\begin{figure}[h]
\begin{center}
\leavevmode
\hbox{%
\epsfxsize=0.6\textwidth
\epsffile{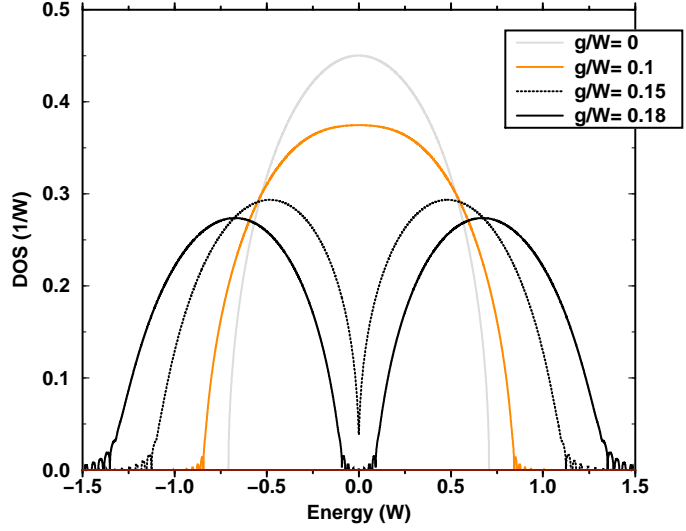}}
\end{center}
\caption{Development of the zero-energy pseudogap and subbands in the spectrum
with increasing electron-phonon coupling $g/W$. Plot is for the (forced)
zero-temperature paramagnetic state with $S=J=\infty$, $h=0$, $n=0.5$ and
$\omega/W=0.05$.\label{fDOSg}}
\end{figure}

\begin{figure}[h]
\begin{center}
\leavevmode
\hbox{%
\epsfxsize=0.6\textwidth
\epsffile{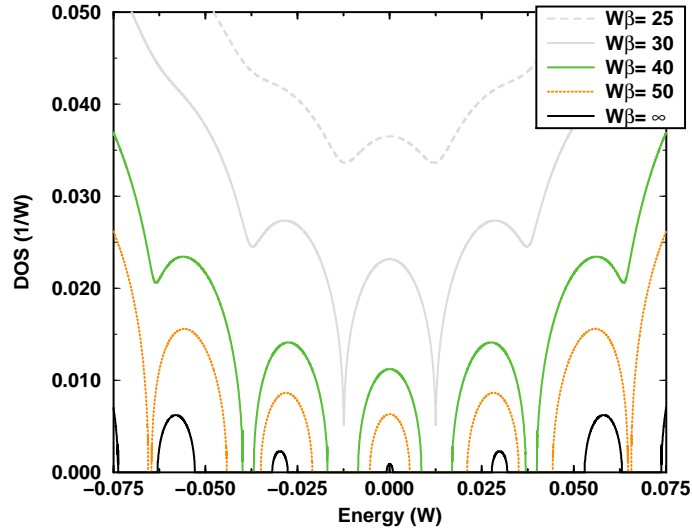}
}
\end{center}
\caption{Evolution of the subbands in the pseudogap (see Fig.\
\ref{fDOSg}) with
temperature. Plot is for the (forced) paramagnetic state with $S=J=\infty$,
$h=0$, $n=0.5$, $\omega/W=0.05$ and $g/W=0.18$.\label{fDOST}}
\end{figure}

The effects of phonon coupling on the (forced) $h=T=0$ paramagnetic state
DOS are shown in Fig.\  \ref{fDOSg}. At $g=0$ we obtain the usual elliptic
band \cite{rUs1}.
As the coupling $g$ is increased the DOS broadens, small subbands are
split off from the band-edges, and a pseudogap appears near the Fermi energy.
At a critical value $g_c$ the DOS splits near zero energy, leaving a small
polaron band in the gap with low weight but very large mass.
Increasing $g$ further causes more bands to be formed 
in the gap, with weights equal to the relevant atomic limit peak weights.
The effect of increasing the temperature $T$ on the DOS in the pseudogap is
shown in Fig.\  \ref{fDOST} for $g=0.18W>g_c$. With increasing
$T$ the DOS at the Fermi surface increases rapidly and the polaron bands are
smeared out. 

For $g>g_c$ the majority of electrons are in fully occupied bands,
and the itinerant electrons lie in a polaron band of very small weight near
zero energy. At $T=0$ this band is equivalent to
the one obtained in the standard strong-coupling theory of the Holstein model
\cite{rLangFirsov}, where one averages the phonons out of the
Hamiltonian $\tilde{H}$ considering only
diagonal electron hopping processes in which
the number of phonons in each state
is conserved. In our approximation this polaron band is damped even at $T=0$
(i.e.\ Im$\Sigma(\mu)\ne$0),
but it would be coherent (barring damping coming from the disordered local
spins, i.e.\ in the saturated ferromagnetic state)
in an approximation which took better account of the
dynamics. In the usual strong-coupling treatment, which treats only the
coherent polaron band, it is found that the DOS $D(\mu)$ at the Fermi surface
decreases with increasing $T$. This is not inconsistent with our finding that
$D(\mu)$ increases with $T$ since our DOS includes all the spectral weight,
both (ideally) coherent and incoherent. For $n\ne 0.5$ the DOS is no longer
symmetric about zero energy; the main lower and upper bands into which the DOS
is split for $g>g_c$ have approximate weights $n$ and $1-n$ respectively.
Although these large features of the spectrum vary considerably with doping
the zero-temperature chemical potential is always confined to the polaron band 
near zero energy, moving from the bottom at $n=0$ to the top at $n=1$ (so that
we obtain a band insulator in these cases).

In our CPA we have no reliable means of calculating the probability
distribution function $P(S^z)$, so to go below $\Tc$ we use the mean-field
approximation for the ferromagnetic Heisenberg model with classical spins and
nearest-neighbour coupling. Since we regard our CPA as an approximation to
DMFT, which is also exact in the infinite-dimensional limit, this simple
approximation may not be unreasonable. We obtain the coupling constant for the
Heisenberg model by matching Curie temperatures. We take this coupling constant
to be temperature-independent, but in a more systematic mapping onto the
Heisenberg model one would expect the coupling constant to vary with
temperature. Note also that in principle the Heisenberg
model's $P(S^z)$ is of a different form to the DE model's \cite{rFurukawaTc}.
One effect of using a mean-field approximation for the magnetisation
will be to obtain the mean-field magnetisation exponent of $1/2$; in three
dimensions we expect the magnetisation to increase more rapidly below $\Tc$,
but note that Schwartz \etal \cite{rSchwartz} find that the magnetisation
exponent in La$_{0.8}$Sr$_{0.2}$MnO$_3$ is 0.45$\pm$0.05.

\begin{figure}[h]
\begin{center}
\leavevmode
\hbox{%
\epsfxsize=0.6\textwidth
\epsffile{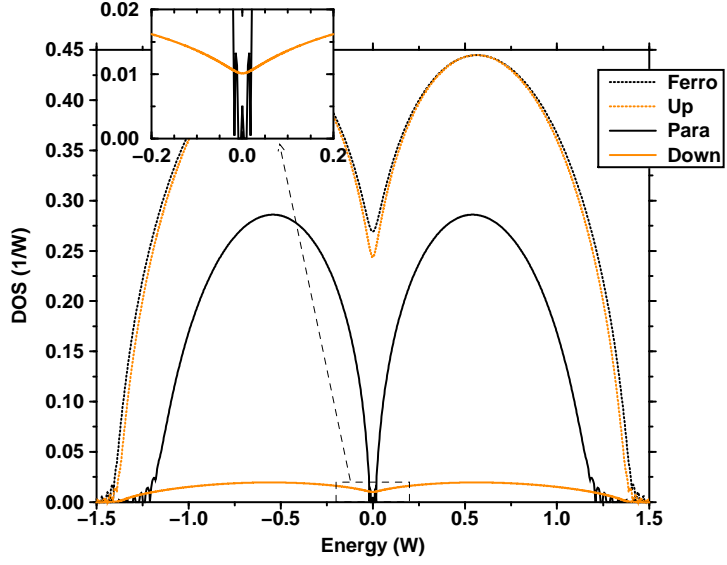}
}
\end{center}
\caption{Comparison of low-temperature spectra for different magnetisations:
saturated ferromagnetism and (forced) paramagnetism for $T=0$ versus
calculated up- and down-spin spectra for $k_{\rm B}T=
0.005W\ll k_{\rm B}\Tc$ where $\las S^z\ras=0.915$. Plot is for 
$S=J=\infty$, $h=0$, $n=0.5$, $\omega/W=0.05$ and $g/W=0.16$.
\label{fDOSbelowTc}}
\end{figure}

We plot the up- and down-spin DOSs for $T=0.005W/k_{\rm B}\ll\Tc$ and
$g=0.16W>g_c$ in
Fig.\  \ref{fDOSbelowTc}, also showing the saturated ferromagnetic and
paramagnetic state DOSs for comparison. The large difference between
${D}_{\rm ferro}(\mu)$ and ${D}_{\rm para}(\mu)$ mean that for
given $T$ we expect the paramagnetic state to have a much higher resistivity
than a magnetised state. Note that there are no separated polaron bands near
$\mu$ in the up- and
down-spin DOSs, even at this low temperature $k_{\rm B}T=0.1\omega$ where
quantum effects might be expected to be important. The transfer of weight to
the up-spin DOS has broadened the polaron bands enough to remove the gaps in
the DOS, and mixing of the down-spins with the up-spins via the Hund coupling
suffices to remove the gaps from the down-spin DOS too. It therefore appears
that the development of magnetisation prevents quantum effects from becoming
important for this coupling strength, at least as far as the DOS is concerned.

\begin{figure}[h]
\begin{center}
\leavevmode
\hbox{%
\epsfxsize=0.6\textwidth
\epsffile{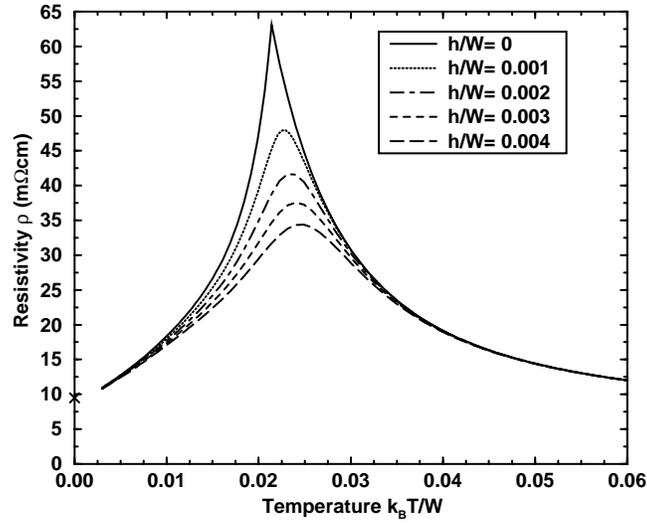}
}
\end{center}
\caption{Resistivity versus temperature for $S=J=\infty$, $n=0.5$,
$\omega/W=0.05$ $g/W=0.16$ and various $h$. The lattice constant $a=5\AA$.
\label{fres}}
\end{figure}

We calculate the resistivity $\rho$
using the formula obtained in Ref.\ \onlinecite{rUs1},
plotting it against temperature in Fig.\  \ref{fres}
for various magnetic fields
$h$. The form of the curve is broadly in agreement with experiment
\cite{rRamirez}, with the
resistivity peak (which occurs at $\Tc$)
of the correct order of magnitude for La$_{0.75}$Ca$_{0.25}$MnO$_3$
(Ref.\ \onlinecite{rSchiffer})
and we find a large negative magnetoresistance near the peak. Note that for
$W\sim$1eV the Curie temperature $\Tc\sim$230K and the magnetic field
$B=0.004W/(\gmub)\sim$20T (for $g\sim 7/2$). The main differences between
Fig.\  \ref{fres} and experiment are the large residual $T=0$ resistivity,
due to the artificial incoherence of the CPA, and the less rapid drop in the
resistivity below $\Tc$ and with $h$, possibly
due to the mean-field form used for the
magnetisation. The rise in the resistivity below $\Tc$ is due to the effects
of the reduced spin polarisation on the DOS. Below $\Tc$ these effects
dominate over the effects on the DOS of thermal smearing, which is responsible
for the fall in the resistivity above $\Tc$.

\begin{figure}[h]
\begin{center}
\leavevmode
\hbox{%
\epsfxsize=0.6\textwidth
\epsffile{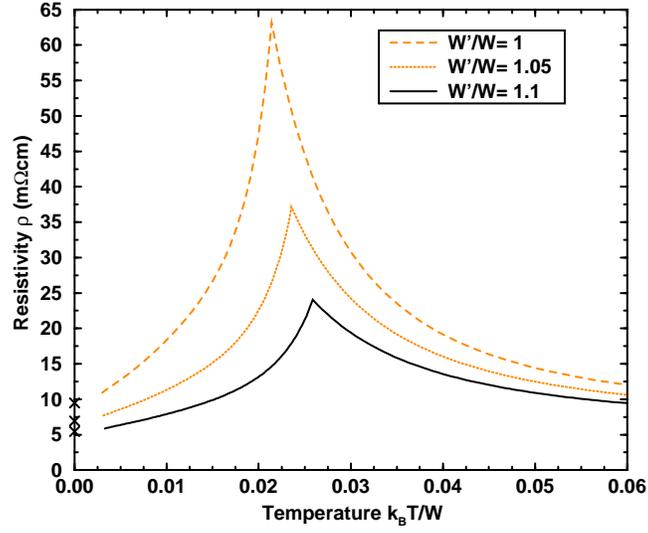}
}
\end{center}
\caption{Effect of pressure (increasing half-bandwidth $W$)
on the resistivity. Plot is for $S=J=\infty$, $n=0.5$, $\omega/W=0.05$
and $g/W=0.16$. The lattice constant $a=5\AA$.
\label{frespress}}
\end{figure}

In Fig.\  \ref{frespress} we show the effect on the resistivity of increasing
hydrostatic pressure, which we model as an increase in the bandwidth. Any
change in the other terms of the Hamiltonian is neglected (note that the
resistivity is proportional to the lattice constant $a$, so a decrease in $a$
will only reinforce the trend observed in Fig.\  \ref{frespress}). The strong
suppression of the peak and the increase in Curie temperature is in agreement
with the measurements of Neumeier \etal \cite{rNeumeier} on
La$_{0.67}$Ca$_{0.33}$MnO$_3$, where a drop in peak resistivity of a factor of
$\sim 2$ is observed when a pressure of 1.62GPa is applied. This change comes
mainly from a decease in the effective coupling constant $g^2/(\omega W)$.

\section{The Holstein model}\label{sHolstein}

We now briefly consider the special case $J=h=0$ where Hamiltonian
(\ref{eHDE}) 
reduces to the Holstein model. In this case CPA equation (\ref{eCPA})
takes the 
form
\begin{equation}
\left[E_{\s}(\en)+K^{\alpha}(\theta,\phi)\right]
\lla \exp\left(\phi b^{\dag}_i\right)\exp\left(\theta b_i\right)
n^{\alpha}_i c_{i\s}\uu\,;c^{\dag}_{i\s}\rra_{\en}
\approx \la \exp\left(\phi b^{\dag}\right)
\exp\left(\theta b\right)
n^{\alpha}_{-\s}\ra,
\end{equation}
and can be solved as in Sec.\  \ref{sGF} to yield
\begin{eqnarray}\nonumber
G_{\s}(\en)=\sum_{r=-\infty}^{\infty}\frac{\I_r
\left(2\l\sqrt{b(\omega)(b(\omega)+1)}\,\right)}
{\exp(\l(2b(\omega)+1))}\left[
\frac{e^{r\beta\omega/2}\las n^-_{-\s}n_{\s}\ras+
e^{-r\beta\omega/2}\las n^-_{-\s}n^-_{\s}\ras}
{E_{\s}(\en)+\lambda\omega+\omega r}\right.\\ \left.
+\frac{e^{r\beta\omega/2}\las n_{-\s}n_{\s}\ras+
e^{-r\beta\omega/2}\las n_{-\s}n^-_{\s}\ras}
{E_{\s}(\en)+3\lambda\omega+\omega r}\right].\label{eHolstein}
\end{eqnarray}
There is now no mixing of up- and down-spins in the problem so our CPA takes
the simple form $G(\en)=G_{\rm AL}(E(\en))$ (where $G_{\rm AL}$ is the atomic
limit Green function)
 which one would guess for a many-body CPA: Eq.\ (\ref{eHolstein})
is just the atomic limit result Eq.\ (\ref{eHolsteinAL}) with
$\en\mapsto E_{\s}(\en)$. The first ($G^{\alpha=-}_{\s}$)
term of Eq.\ (\ref{eHolstein}) corresponds to
polaron bands near energy $-\lambda\omega$, and the second
($G^{\alpha=+}_{\s}$) is the
bipolaronic term, corresponding to bands near $-3\lambda\omega$. Our
approximation's reliance on the atomic limit means that all bipolaron coupling
takes place on-site.

In principle we can use the spectral theorem to determine all weights
self-consistently in terms of the Green functions $G^{\alpha}_{\s}$, but for
low temperature and strong coupling $g$ most electrons are bound as
bipolarons and we may set $\las n_{\u}n_{\dd}\ras\approx\las n_{\s}\ras=n/2$. 
The groundstate of the Holstein model is actually believed to be either
superconducting (away from half-filling and at strong coupling) or a charge
density wave (near half-filling and at weak coupling) \cite{rCiuchi2}. 
However, determining the weights self-consistently near the homogeneous state
we do not find a (second-order) transition to a charge density wave state.
This is
reminiscent of the CPA for the Hubbard model, where no transition to
ferromagnetism or antiferromagnetism exists. We are also unable to consider
superconductivity within our approximation.

Note that the true CPA/DMFT result in the
empty-band limit, first obtained by Sumi \cite{rSumi} using the CPA and
rederived by Ciuchi \etal \cite{rCiuchi} using DMFT, is
\begin{equation}
G(\en)= (1-e^{-\beta\omega})\sum_{n=0}^{\infty}\frac{e^{-n\beta\omega}}
{E(\en)-A_n(\en)-B_n(\en)}\label{etrueCPA}
\end{equation}
where $A_n$ is the finite continued fraction
\begin{mathletters}
\begin{equation}
A_n(\en)=\frac{ng^2}{E(\en+\omega)-\frac{(n-1)g^2}{E(\en+2\omega)-
\frac{(n-2)g^2}{^{\cdot}\cdot_{\cdot}-\frac{g^2}{E(\en+n\omega)}}}}
\label{etrueCPA1}
\end{equation}
and $B_n$ is the infinite continued fraction
\begin{equation}
B_n(\en)=\frac{(n+1)g^2}{E(\en-\omega)-\frac{(n+2)g^2}{E(\en-2\omega)
-\frac{(n+3)g^2}{E(\en-3\omega)-\cdots}}}.\label{etrueCPA2}
\end{equation}
\end{mathletters}
As mentioned in Sec.\  \ref{sGF} our result in the empty-band limit is only
equivalent to the true one-electron CPA result if the approximation
$E(\en+r\omega)\approx E(\en)+r\omega$ is made in equations
(\ref{etrueCPA1}) and (\ref{etrueCPA2}).


\section{Summary}\label{sSummary}

In this paper we have extended our many-body CPA, developed in references
\onlinecite{rUs1} and \onlinecite{rUs2}
for the DE model, to study the Holstein-DE model, which we regard
as a simple model for CMR materials. We were interested in effects due to the
quantisation of the phonons. Our CPA has the advantage over DMFT of being
analytically relatively simple, although necessarily cruder, and over
variational Lang-Firsov approaches of being able to study the whole of the
spectrum, not just the low-energy coherent polaron band. We solved the
Holstein-DE model exactly in the atomic limit in which
the CPA becomes exact and solved the CPA equations in the strong
Hund-coupling limit $J\rightarrow\infty$. Using a DMFT result for the
local spin polarisation in terms of electronic Green functions we
obtained an equation for the Curie temperature $\Tc$.
For intermediate electron-phonon coupling strength we obtained
reasonable agreement with experiment for most calculated quantities, including
the Curie temperature and resistivity. It appears however that for this
range of coupling the development
of magnetisation below $\Tc$ prevents the quantisation of the phonons from
affecting the DOS near the Fermi surface even at low temperatures.

\section*{Acknowledgements}
I am grateful to DM Edwards for helpful discussions and to the UK
Engineering and Physical Sciences Research Council (EPSRC) for financial
support.

\appendix
\section*{}

We now solve Eq.\ (\ref{ePDE1})
for $S_{\u}$ using the method of characteristics.
For compactness of notation we define the operator $U=
\left[(S+1+S^z)n^-_{\dd}+S^-\s^+\right]/(2S+1)$.
In the $(\rho,\Theta,\Phi)$ system of variables Eq.\ (\ref{ePDE1})
takes the form
\begin{eqnarray}\nonumber
\left[\frac{E^h_{\u}(\en)-E^h_{\dd}(\en)}{2S+1}\right]
\frac{\partial S_{\u}}{\partial\rho}+\frac{\partial S_{\u}}{\partial\Theta}
=\la e^{\rho S^z}e^{\phi(\Theta)b^{\dag}}e^{\theta(\Theta,\Phi)b}\,
U\ra\hskip 2in\\
-\left[\frac{g^2}{\omega}+g\Phi e^{-\omega\Theta}+
\frac{(1+S)E^h_{\u}(\en)+SE^h_{\dd}(\en)}{2S+1}\right]
S_{\u}
\end{eqnarray}
where $\phi(\Theta)=e^{\omega\Theta}-g/\omega$ and
$\theta(\Theta,\Phi)=\Phi e^{-\omega\Theta}+g/\omega$. The characteristic
equations are
\begin{mathletters}
\begin{eqnarray}
\frac{\d\rho}{\d s}&=&\left[\frac{E^h_{\u}(\en)-E^h_{\dd}(\en)}{2S+1}\right],
\hskip 1in
\frac{\d\Theta}{\d s}=1\\
\frac{\d S_{\u}}{\d s}&=&\la e^{\rho S^z}e^{\phi(\Theta)b^{\dag}}
e^{\theta(\Theta,\Phi)b}\,U\ra-
\left[\frac{g^2}{\omega}+g\Phi e^{-\omega\Theta}
+\frac{(1+S)E^h_{\u}(\en)+SE^h_{\dd}(\en)}{2S+1}\right]
S_{\u}.\label{echar3}
\end{eqnarray}
\end{mathletters}
The first two are solved immediately as
\begin{equation}
\rho=\rho_0+\left[\frac{E^h_{\u}(\en)-E^h_{\dd}(\en)}{2S+1}\right]s,
\hskip 1in \Theta=s
\end{equation}
where $\rho_0$ is an arbitrary constant and we set the constant in the
$\Theta$ equation to zero without loss of generality.
These solutions are substituted into Eq.\ (\ref{echar3}),
which may then be written as
\begin{eqnarray}\nonumber
\frac{\d}{\d s}
\left[e^{\left(
\left[(1+S)E^h_{\u}(\en)+SE^h_{\dd}(\en)\right]/(2S+1)+
g^2/\omega
\right)s
-\frac{g}{\omega}\Phi e^{-\omega s}}S_{\u}\right]=
e^{\left(\left[(1+S)E^h_{\u}(\en)+SE^h_{\dd}(\en)\right]/(2S+1)+
g^2/\omega\right)s}
\\ \times\la e^{
\left(\rho_0+s\left[E^h_{\u}-E^h_{\dd}\right]/(2S+1)\right)S^z}
e^{-\frac{g}{\omega}b^{\dag}}
e^{e^{\omega s}b^{\dag}}
e^{\Phi e^{-\omega s}\left(b-\frac{g}{\omega}\right)}
e^{\frac{g}{\omega}b}\,U\ra.\label{echar3b}
\end{eqnarray}
We expand $\exp(e^{\omega s}b^{\dag})$ and
$\exp(\Phi e^{-\omega s}(b-g/\omega))$ in
Eq.\ (\ref{echar3b}) as series and integrate to find
\begin{eqnarray}\nonumber
S_{\u}\exp\left\{\left(\frac{(1+S)E^h_{\u}+SE^h_{\dd}}{2S+1}+
\frac{g^2}{\omega}\right)s-\frac{g}{\omega}\Phi e^{-\omega s}\right\}=S_{\u 0}
+\sum_{m,n=0}^{\infty}\frac{\Phi^n}{m!n!}\la e^{\rho_0 S^z}
e^{-\frac{g}{\omega}b^{\dag}}\left(b^{\dag}\right)^m\right.\\ 
\left. \times\frac{(2S+1)e^
{\left([(1+S+S^z)E^h_{\u}+(S-S^z)E^h_{\dd}]/(2S+1)+g^2/\omega
+(m-m)\omega\right)s}}{
(1+S+S^z)E^h_{\u}+(S-S^z)E^h_{\dd}+(2S+1)(g^2/\omega
+(m-m)\omega)}\left(b-\frac{g}{\omega}\right)^n e^{\frac{g}{\omega}b}U\ra.
\end{eqnarray}
We then write the characteristics as intersections of the surfaces
$S_{\u 0}=S_{\u 0}(\rho,\Theta,\Phi)$ and $\rho_0=\rho_0(\rho,\Theta,\Phi)$,
and the general solution of Eq.\ (\ref{ePDE1}) is of the form
$S_{\u 0}(\rho,\Theta,\Phi)=F(\rho_0(\rho,\Theta,\Phi))$ where $F$ is an
arbitrary function. Rearranging we obtain
\begin{eqnarray}\nonumber
S_{\u}=\exp\left(\frac{g}{\omega}\Phi e^{-\omega\Theta}\right)\left\{
\exp\left[-\left(\frac{(1+S)E^h_{\u}+SE^h_{\dd}}{2S+1}+\l\omega\right)
\Theta\right]F\left(\rho-\frac{E^h_{\u}-E^h_{\dd}}{2S+1}\Theta\right)\right.
\hskip 0.25in\\
\left.+\sum_{m,n=0}^{\infty}\frac{\Phi^n}{m!n!}
e^{(m-n)\omega\Theta}\la\frac{
(2S+1)e^{\rho S^z}e^{-g/\omega\, b^{\dag}}(b^{\dag})^m
(b-g/\omega)^n e^{g/\omega\, b}U}
{(S+1+S^z)E^h_{\u}+(S-S^z)E^h_{\dd}+(2S+1)(\l\omega+(m-n)\omega)}\ra
\right\}.\label{eextra}
\end{eqnarray}
Now from definition Eq.\ (\ref{edefS}) of
$S_{\u}$ it may be seen that $S_{\u}$ is of the form
\begin{equation}
S_{\u}\left(\rho+\frac{E^h_{\u}-E^h_{\dd}}{2S+1}\Theta,\Theta,\Phi\right)=
\sum_{m=-S}^{S}\sum_{n=-\infty}^{\infty}a_{mn}(\rho,\Phi)
\exp\left[\left(m\frac{E^h_{\u}-E^h_{\dd}}{2S+1}+n\omega\right)\Theta\right].
\end{equation}
The final term of Eq.\ (\ref{eextra})
is compatible with this form but the term
proportional to $F$ is not, so we must have $F\equiv 0$. Finally, in our
original $(\rho,\theta,\phi)$ system of variables
\begin{eqnarray}\nonumber
S_{\u}(\rho,\theta,\phi)=
\exp\left(\frac{g}{\omega}\left(\theta-\frac{g}{\omega}\right)\right)
\sum_{m,n=0}^{\infty}\frac{(\theta-g/\omega)^n(\phi+g/\omega)^m}{m!\,n!}
\hskip 1.5in\\ \times
\la\frac{(2S+1)e^{\rho S^z}e^{-g/\omega\,b^{\dag}}\left(b^{\dag}\right)^m
\left(b-g/\omega\right)^n e^{g/\omega\,b}}
{(1+S+S^z)E^h_{\u}+(S-S^z)E^h_{\dd}+(2S+1)(g^2/\omega+(m-n)\omega)}U\ra.
\end{eqnarray}

\end{document}